\documentclass[proceedings,paper]{JHEP3}
\PrHEP{PrHEP hep2001}                   
\conference{International Europhysics Conference on HEP}
\usepackage{epsfig}
\def\AGUT{{}\;\;\raisebox{.9ex}{$\times$}\raisebox{-.5ex}%
{$\!\!\!\!\!\!\!\!\sss i=1,2,3$} \,(SMG_i \times U(1)_{\sss B-L,i})}

\def\ie{\hbox{\it i.e.}{}}
\def\eg{\hbox{\it e.g.}{}}

\def\nn{\hspace{2mm}}
\def\sss{\scriptscriptstyle}
\def\MeV{\mbox{\rm MeV}}
\def\GeV{\mbox{\rm GeV}}

\def\eV{\mbox{\rm eV}}

\def\sVEV#1{\left\langle #1\right\rangle}

\def\abs#1{\left| #1\right|} 

\def\sleq{\raisebox{-.6ex}{${\textstyle\stackrel{<}{\sim}}$}}

{\title{A Non-SUSY Model for Neutrino Oscillation, Baryogenesis 
and Neutrinoless Double Beta Decay}}
\author{Holger Bech Nielsen\thanks{Talk presented at the International 
Europhysics Conference on High Energy Physics (EPS HEP 2001), 
Budapest, Hungary, July 12-18, 2001.}{},~Yasutaka Takanishi\thanks{Talk  
presented at the Les Houches EuroConference on Neutrino masses 
and mixings, Les Houches, France, June 18-22, 2001.}\\ The Niels Bohr  
Institute, Blegdamsvej 17, \\ DK-2100 Copenhagen {\O}, 
Denmark\\ E-mails: \email{hbech@nbi.dk},~~\email{yasutaka@nbi.dk}}

\abstract{We fitted the neutrino oscillations, 
baryogenesis and neutrinoless double beta decay using Anti-GUT model.}

\keywords{Fermion Masses, Neutrino Oscillations, Baryogenesis, 
Neutrinoless beta decay}

\preprint{\hepph{0110125}~~~NBI-HE-01-10}

\begin{document}
\vspace*{-.23cm}
\section{Introduction}
\vspace*{-.23cm}

\indent

The Anti-GUT model \cite{NT} is based on a large gauge group,
$\AGUT$, where $i$ denoted as the generation numbers, $\ie$, each family
has its own Standard Model (SM) gauge group with additional gauged 
Baryon number minus Lepton number, $B-L$. The fermions are the $45$ 
SM Weyl ones plus three right-handed neutrinos. 

This gauge group characterised as the largest subgroup of the $U(48)$
transforming the Weyl fermions without anomalies, neither mixed nor 
gauge ones, and not unifying 
irreducible representation in SM. It is broken down by the six Higgs
fields of Table $1$ to the SM at the weak scale. The Weinberg-Salam
Higgs field denoted as $\phi_{WS}$ is also in the table.

The right-handed neutrinos are mass-protected by the total gauge group
$B-L$ (diagonal subgroup of the three family $B-L$'s) and obtain the
see-saw scale mass given by the vacuum expectation value (VEV) of the
Higgs field, $\phi_{B-L}$, which breaks $B-L$ quantum charge by 
two units.

The remaining Higgs fields have VEV one or two order magnitude
under the Planck scale, except the Higgs field, $S$, the VEV of
which is almost of the order of the Planck scale.

\newpage
\vspace*{-2mm}  
\TABULAR[ht]{ccccccc}{\hline\hline & $SMG_1$& $SMG_2$ & $SMG_3$ & $U_{\sss B-L,1}$ & $U_{\sss B-L,2}$ & $U_{\sss B-L,3}$ \\ \hline
$\phi_{\sss WS}$ & $\frac{1}{6}$ & $\frac{1}{2}$ & $-\frac{1}{6}$ & $-\frac{2}{3}$ & $1$ & $-\frac{1}{3}$ \\
$S$ & $\frac{1}{6}$ & $-\frac{1}{6}$ & $0$ & $-\frac{2}{3}$ & $\frac{2}{3}$ & $0$ \\        
$W$ & $-\frac{1}{6}$ & $-\frac{1}{3}$ & $\frac{1}{2}$ & $\frac{2}{3}$ & $-1$ & $\frac{1}{3}$ \\
$\xi$ & $\frac{1}{3}$ & $-\frac{1}{3}$ & $0$ & $-\frac{1}{3}$ & $\frac{1}{3}$ & $0$ \\
$T$ & $0$ & $-\frac{1}{6}$ & $\frac{1}{6}$ & $0$ & $0$ & $0$ \\
$\chi$ & $0$ & $0$ & $0$ & $0$ & $-1$ & $1$ \\
$\phi_{\sss B-L}$ & $-\frac{1}{6}$ & $\frac{1}{6}$ & $0$ & $\frac{2}{3}$ &$-\frac{2}{3}$ & $2$\\ \hline\hline}{All $U(1)$ quantum charges in Anti-GUT model. $SMG_i$ denote the SM gauge groups.}
In the table we have only Abelian quantum numbers of the Higgs fields;
it is understood that non-Abelian representations are taken to be 
smallest ones obeying the following rules:
\begin{equation}
  \label{eq:SMrule}
  d_i/2+t_i/3+y_i/2=0\;({\rm mod}\;1)\nn.
\end{equation}
Here $t_i$ is the triality being $1$ for $\underline{3}$, 
$-1$ for $\overline{\underline{3}}$ and $0$ for $\underline{1}$ or 
$\underline{8}$, respectively. The duality $d_i$ is $0$ for integer 
spin $SU(2)_i$ representations while $1$ for half-integer $SU(2)_i$ spin.

Each generation gauge group is subgroup of $SO(10)$,
$\ie$, $\AGUT\subset SO(10)^3$, and the fermion representations
could be extended to $SO(10)$ spinor representation but our Higgs field
representations could not be enrolled as $SO(10)$ representations.

The philosophy of ``dull'' Model\footnote{We thank to C.~Jarlskog for 
question concerning this philosophy.} on which the present model 
a long way is based means that we seek a model beyond SM which 
as similar to SM as possible - only deviating from the latter when 
phenomenologically required. We could for example hold that having 
several repetitions of SM gauge field systems, one for each 
generation, is rather 
little new compared to the SM. It might be phenomenologically 
required to have more gauge fields to make families have different 
masses order of magnitude-wise. 

To make as weak assumptions as possible we allow everything to go on at Planck
scale, actually we assume that all Yukawa coupling constants
are order one complex numbers, so that we can treat them as random 
numbers. Therefore, we can predict from this model everything 
{\em only} with order of magnitude accuracy.

\vspace*{-.23cm}
\section{Mass matrices and results for masses and mixing angles}
\vspace*{-.23cm}
\indent

Effective Yukawa couplings appear due to the transition
of using our several Higgs fields~\cite{FN}; we have investigated
numerical corrections~\cite{FNS} due to the different orders of Higgs fields 
VEVs being attached to the exchange chain of propagators -- 
left-right transition for the Weyl particles. In the neutrino sector
according to the see-saw mechanism we have to calculate Dirac- and
Majorana-mass matrices:
\begin{equation}
  \label{eq:meff}
  M_{\rm eff} \! \approx \! M^D_\nu\,M_R^{-1}\,(M^D_\nu)^T\nn.
\end{equation}
Here we present all mass matrices:
\begin{eqnarray*}
M_U &\simeq& \frac{\sVEV{\phi_{\sss\rm WS}}}{\sqrt{2}}\hspace{-0.1cm}
\left(\begin{array}{ccc}
        6\sqrt{35} SWT^2\xi^2
        & 6\sqrt{10} SWT^2\xi
        & 6\sqrt{35} S^2W^2T\xi \\
        12\sqrt{105}S^{2}WT^2\xi^3
        & 2\sqrt{3}WT^2
        & 2\sqrt{15}SW^2T \\
        2\sqrt{35}S^{3}\xi^3
        & \sqrt{2}S
        & \sqrt{6}WT
\end{array} \right)\label{M_U} \nn,\\
M_D &\simeq& \frac{\sVEV{\phi_{\sss\rm WS}}}{\sqrt{2}}\hspace{-0.1cm}
\left(\begin{array}{ccc}
        6\sqrt{35}SWT^2\xi^2
      & 6\sqrt{10}SWT^2\xi
      & 2\sqrt{105}S^2T^3\xi \\
        2\sqrt{15}WT^2\xi
      & 2\sqrt{3}WT^2
      & 2\sqrt{5}ST^3 \\
        6\sqrt{210}SW^2T^4\xi
      & 2\sqrt{210}SW^2T^4
      & \sqrt{6}WT \nn,\\
\end{array} \right) \label{M_D} \\
M_E &\simeq& \frac{\sVEV{\phi_{\sss\rm WS}}}{\sqrt{2}}\hspace{-0.2cm}
\left(\begin{array}{ccc}
       6\sqrt{35}SWT^2\xi^2
  & 60\sqrt{14}S^{3}WT^2\xi^3
  & 60\sqrt{2002}S^{3}WT^4\xi^3\chi\\
    6\sqrt{30030}S^4WT^2\xi^5
  & 2\sqrt{3}WT^2
  & \sqrt{210}WT^4\chi\\
    36\sqrt{3233230}S^7W^2T^4\xi^5
  & 30\sqrt{14}S^3W^2T^4
  & \sqrt{6}WT\end{array} \right) \label{M_E}\nn,\\
M^D_\nu &\simeq&\frac{\sVEV{\phi_{\sss\rm WS}}}{\sqrt{2}}
\left(\begin{array}{ccc}6 \sqrt{35}\,S\,W\,T^2\,\xi^2 & 60 
\sqrt{14}\,S^3\,W\,T^2\,\xi^3
& 60\sqrt{154}\,S^3\,W\,T^2\,\xi^3\,\chi\\
6 \sqrt{35}\,S^2\,W\,T^2\,\xi & 2\sqrt{3}\,W\,T^2 & 2\sqrt{15}\,W\,T^2\,\chi \\
6\sqrt{70}\,S^2\,W\,T\,\xi\,\chi & 2\sqrt{6} W\,T\chi & \sqrt{6} W\,T
\end{array} \right) \nn,\\
M_R &\simeq&\sVEV{\phi_{\sss\rm B-L}}
\left (\begin{array}{ccc}2\sqrt{210} S^{3}\chi^2\xi^2 & 
\sqrt{15}S\chi^2\xi & \sqrt{6}S\chi \xi  \\
\sqrt{15} S\chi^2\xi & \sqrt{6} S\chi^2 & \sqrt{\frac{3}{2}} S\chi \\
\sqrt{6} S\chi\xi & \sqrt{\frac{3}{2}} S\chi & S
\end{array} \right ) \nn,
\end{eqnarray*}

\TABULAR[hb!]{c|c|c}{\hline\hline 
& Fitted & Experimental \\ \hline \hline
$m_u$ & $3.1$~$\MeV$ & $4$~$\MeV$ \\
$m_d$ & $6.6$~$\MeV$ & $9$~$\MeV$ \\
$m_e$ & $0.76$~$\MeV$ & $0.5$~$\MeV$ \\
$m_c$ & $1.29$~$\GeV$ & $1.4$~$\GeV$ \\
$m_s$ & $390$~$\MeV$ & $200$~$\MeV$ \\
$m_{\mu}$ & $85$~$\MeV$ & $105$~$\MeV$ \\
$M_t$ & $179$~$\GeV$ & $180$~$\GeV$ \\
$m_b$ & $7.8$~$\GeV$ & $6.3$~$\GeV$ \\
$m_{\tau}$ & $1.29$~$\GeV$ & $1.78$~$\GeV$ \\
$V_{us}$ & $0.21$ & $0.22$ \\
$V_{cb}$ & $0.023$ & $0.041$ \\
$V_{ub}$ & $0.0050$ & $0.0035$ \\
$J_{\sss CP}$ & $1.04\times10^{-5}$ & $2\!-\!3.5\times10^{-5}$\\ 
\hline \hline}{A fit including averaging over ${\cal O}(1)$ factors.
All quark masses are running masses at
$1~\GeV$ except the top quark mass which is the pole mass.}
\noindent
where $M_U$ up-type, $M_D$ down-type, $M_E$ charged lepton,
$M^D_\nu$ Dirac-neutrino $M_R$ Majorana-neutrino mass matrix, 
respectively. Each matrix elements are understood to be further
multiplied by order one random numbers.

We present the best fit using following VEVs for charged fermion 
quantities in Table $2$ and neutrino quantities in Table $3$:
\begin{eqnarray}
&&\sVEV{\phi_{\sss WS}}=\frac{246}{\sqrt{2}}~\GeV\nn, 
\nn\sVEV{\phi_{\sss B-L}}=2.74\times10^{12}~\GeV\nn, \nonumber\\
&&\sVEV{S}=0.721\nn,\nn\sVEV{W}=0.0945\nn,\nn\sVEV{T}=0.0522\nn,\nonumber\\
&&\sVEV{\xi}=0.0331\nn,\nn\sVEV{\chi}=0.0345\nn,
\end{eqnarray} where the VEVs, except the Weinberg-Salam
Higgs and $\sVEV{\phi_{\sss B-L}}$, are presented in the Planck unit.

Under the impression of the combination of Day Night effect at 
Super-Kamiokande and the first results from SNO we are able to 
get new version of this type of model
concerning the first to second transition - replacing the Higgs 
fields, $\xi$ and $S$, by new Higgs fields, $\omega$ and $\rho$, which
have quantum numbers 
\begin{equation}
  \omega=(\frac{1}{6},-\frac{1}{6},0,0,0,0)\nn {\rm and} \nn
\rho=(0,0,0,-\frac{1}{3},\frac{1}{3},0)\nn. 
\end{equation}

\newpage

\TABULAR[h!!l]{c|c|c}{\hline\hline
& Fitted & Experimental \\ 
\hline \hline
$\frac{\Delta m_{\odot}^2}{\Delta m_{\rm atm}^2}$ & $5.8{+30\atop-5}\times10^{-3}$ & $ 1.5 {+1.5\atop-0.7}\times10^{-3}$ \\
$\tan^2\theta_{\odot}$ & $ 8.3{+21\atop-6}\,\times10^{-4}$ & $(0.33-2.5)\times 10^{-3}$ \\
$\tan^2\theta_{e3}$ & $4.3{+11\atop-3}\,\times10^{-4}$ & $ 2.6\times10^{-2}$ \\
$\tan^2\theta_{\rm atm}$ & $0.97 {+2.5\atop-0.7}$ & $0.43 - 1.0$ \\
\hline\hline}{The numerical results of the ratio of mass squared 
differences and solar, atmospheric and CHOOZ mixing angles.}

\noindent
They fit large mixing angle MSW solution~\cite{talks}. 

\section{Baryogenesis via lepton \\ number violation}
\indent

In the models having see-saw neutrinos~\cite{FY} we get
at first due to $B-L$ violation ($\sVEV{\phi_{B-L}}$ 
in our case), out-of-equilibrium due to the masses of
right-handed neutrinos and $CP$ violation 
an excess of $B-L$. After this time (see-saw era)
$B-L$ is conserved as an ``accidental'' symmetry 
in the SM, even at temperatures so high that sphalerons allows
violation of $B$ and $L$ separately. 

The $CP$ violation is parameterised by $\epsilon_i$ 
in the decay of the $i$th flavour right-handed neutrino~\cite{buchmuller}:
\begin{equation}
 \epsilon_{i}\equiv{\Gamma_{N_{R_i}\ell}-\Gamma_{N_{R_i}\bar\ell}
\over \Gamma_{N_{R_i}\ell}+ \Gamma_{N_{R_i}\bar\ell}} = \frac{\sum_{j\not= i} 
{\rm Im}[((M_\nu^D)^{\dagger} M_\nu^D)^2_{ji}] [\,f(\frac{M_j^2}{M_i^2} ) + g( \frac{M_j^2}{M_i^2})\,]}{4\pi \sVEV{\phi_{\sss WS}}^2 
((M_\nu^D)^{\dagger} M_\nu^D)_{ii}}\nn, \nonumber
\end{equation}
where $f$ comes from the one-loop vertex contribution and
$g$ comes from the self-energy contribution, which
can be calculated in perturbation theory if Majorana masses
satisfy the condition, $\abs{M_i-M_j}\gg\abs{\Gamma_i-\Gamma_j}$:
\begin{equation}
f(x) =\sqrt{x} \left[1-(1+x) \ln \frac{1+x}{x}\right]\nn,\nn\nn 
g(x) = \frac{\sqrt{x}}{1-x} \nn.
\end{equation}

After creation of $B-L$ asymmetry there can be significant 
wash-out; for it we should use the quantities:
\begin{equation}
K_i\equiv\frac{\Gamma_i}{2 H} \,\Big|_{ T=M_{i} } = \frac{M_{\rm Planck}}
{1.66 \sVEV{\phi_{\sss WS}}^2  8 \pi g_{*\,i}^{1/2}}
\frac{((M_\nu^D)^{\dagger} M_\nu^D)_{ii}}{M_{i}} \qquad
(i=1, 2, 3)\nn, 
\end{equation}
where $\Gamma_i$ is the width of the flavour $i$ Majorana neutrino,
$M_i$ is its mass and $g_{*\,i}$ is the number of the degree of freedom
at temperature $M_i$ and in non-SUSY case approximately $100$. 
The numerical results of our best fitting case (Table $2$ and $3$) gives
\begin{eqnarray}
\abs{\epsilon_3} &=& 6.8 \times 10^{-9}\nn,  \qquad K_3 =1.06 \\
\abs{\epsilon_2} &=& 6.0 \times 10^{-9}\nn,  \qquad K_2 =4.29 \\
\abs{\epsilon_1} &=& 4.8 \times 10^{-10}\nn, \quad\,\,  K_1 =19.8\nn.
\end{eqnarray}
Furthermore, we need the correction from the obtained $K_i$ --
dilution factors -- containing the effect of the sphaleron processes
being given in various ranges of $K_i$ as:
\begin{eqnarray}
10 \sleq\, K_i \,\sleq 10^6:\qquad\kappa_i&=&-\frac{0.3}{K_i(\ln K_i)^{\frac{3}{5}}}\\
0 \sleq\, K_i \,\sleq 10:\qquad\kappa_i&=& -\frac{1}{2\, \sqrt{{K_i}^2 + 9}}
\end{eqnarray}

Using the approximation that there is no exchange between the 
different ``flavour $B-L$'' quantum numbers produced by the 
three different $\nu_{R_i}$'s so that we use only dilution 
with $K_i$ for the $\nu_{R_i}$ produced $B-L$ we got from 
the abovementioned quantities the Baryogenesis via lepton 
number~\cite{NTbary}:
\begin{equation}
\label{Y_B}
   Y_B = \sum_{i=1}^{3}\kappa_i\, \frac{\epsilon_i}{g_{*\,i}}= 1.5{+5.8\atop-1.2}\times10^{-11}\nn.
\end{equation}

That meant that we ignored $\eg$ that resonance scattering via 
the lightest see-saw neutrino as the resonance could 
contribute to the $B-L$ wash-out in the flavour combination produced by 
the decay of $\nu_{R_3}$. Really since the 
couplings of the three see-saw neutrinos are not orthogonal 
in flavour-space this hypothesis is not valid\footnote{We wish to 
thank L.~Covi and M.~Hirsch for useful discussions and comments 
on this problem.} and we might rather crudely estimate an 
effective $K_3$ to be used for the $\nu_{R_3}$ decay products 
as an average of the three $K_i$'s. This would lead to a decrease 
of our prediction of eq.~(\ref{Y_B}) by a factor of the order $6$:
\begin{equation}
   Y_B \approx 2 {+10\atop-1.7}\times10^{-12}\nn.
\end{equation}

Note that the version~\cite{talks} which predicts large mixing 
angle MSW gives {\em not unexpectedly} bigger Baryon number 
production thus improving agreement with experimental date.

\vspace*{-.23cm}
\section{Neutrinoless double beta decay}
\vspace*{-.23cm}
\indent

{}From the fitted quantities, namely, neutrino mass and their
mixing angles we can calculate so-called ``effective Majorana 
mass parameter'' being defined by
\begin{equation}
\abs{\sVEV{m}} \equiv \sum_{i=1}^{3} U_{e i}^2 \,m_i = 5.9{+5.3\atop-2.8}\times10^{-5}~\eV
\end{equation}
where $m_i$ is the mass of the Majorana neutrino $\nu_i$ 
and $U_{e i}$ are the elements of the MNS neutrino mixing 
matrix. The result satisfies the recent experimental limits. 
{}Really, it is clear that a model, which 
predicts small mixing angle MSW for solar neutrino puzzle
and neutrino mass spectra being hierarchical pattern, obeys
the limit of experiments of neutrinoless double beta decay.



\end{document}